\documentstyle[aps,prl,twocolumn,psfig]{revtex}

	\def\al{{\alpha}}
	\def\be{{\beta}}
	\def\bfB{{\bf B}}
	\def\bfS{{\bf S}}
	\def\bfj{{\bf j}}
	\def\bfn{{\bf n}}
	\def\bfv{{\bf v}}
	\def\bfx{{\bf x}}
	\def\const{{\rm const}}
	\def\de{{\delta}}
	\def\ep{{\epsilon}}
	\def\ga{{\gamma}}
	\def\om{{\omega}}
	\def\ph{{\phi}}
	\def\re#1{(\ref{#1})}
	\def\rh{{\rho}}
	\def\ucsd{Department of Physics\\
		University of California at San Diego\\
		La Jolla, CA 92093-0319}
	\newcommand{\bfell}{{\mbox{\boldmath $\bf\ell$}}}
	\newcommand{\bfet}{{\mbox{\boldmath $\bf\eta$}}}
	\newcommand{\bfna}{{\mbox{\boldmath $\bf\nabla$}}}
	\newcommand{\bfom}{{\mbox{\boldmath $\bf\omega$}}}
	\newcommand{\bfxi}{{\mbox{\boldmath $\bf\xi$}}}
	\newcommand{\bfze}{{\mbox{\boldmath $\bf\zeta$}}}
	\newcommand{\p}{{\partial}}

\begin{document}
\bibliographystyle{prsty}

\title{\bf Nonlinear hydrodynamic stability}
\author{M.~B. Isichenko\\ \ucsd}
\date{July 28, 1997}
\maketitle

\begin{abstract}
The variational principle of V.~I. Arnold [J.\ Appl.\ Math.\ Mech.\
Vol.~29, P.~1002 (1965)] is extended to the general conservative
inhomogeneous, compressible, and conducting fluid.  The concept of
iso-vortical flows is generalized to an ``invariant foliation'' of the
phase-space.  The foliation, which may or may not correspond to
explicit conservation laws, is derived from the equations of motion
and used for Lyapunov stability.  A nonlinear three-dimensional
(magneto-) hydrodynamic stability criterion is formulated.
\end{abstract}


The standard approach to hydrodynamic stability involves linearization
about an equilibrium flow in order to solve for eigenfrequencies
\cite{Chandrasekhar61,Lighthill78} or establish a Lyapunov stability
criterion for the linearized system \cite{Berge97}.  A variety of
linear variational principles were developed for both neutral fluids
\cite{Lynden-Bell-Ostriker-67} and magnetohydrodynamics (MHD)
\cite{BFKK58,Frieman-Rotenberg-60,Hameiri-Holties-94}, in which the
stability criterion is expressed in terms of a positive definite
quadratic form.  It is well known that the linearized stability does
not guarantee the true (Lyapunov) stability, such as in the toy system
$du/dt=u^2$, whose equilibrium $u=0$ is linearly stable.

Nonlinear stability is guaranteed by the presence of an integral of
motion, for example, the energy $H$, which assumes a non-degenerate
extremum (a minimum or a maximum) subject to the conservation of any
other integrals of motion, for example, Casimir invariants
\cite{HMRW85}. The possibility to write explicitly a full infinite set
of integrals is mostly limited to two-dimensional systems.  By {\em
explicit\/} we mean an integral of motion which can be written in
terms of the physical fields of velocity, density, etc., in a way
which does not require the solution of the equations of the motion.
In three dimensions, such integrals are scarce.  For example, the
Euler equation
\begin{equation}
\p_t\bfom=\bfna\times(\bfv\times\bfom),\quad\bfom=\bfna\times\bfv,
\quad\bfna\cdot\bfv=0,
\label{euler}
\end{equation}
conserves explicitly only the energy $H$ and the helicity $I$:
\begin{equation}
H=\int\frac{\bfv^2}2,\quad 
I=\int\bfv\cdot\bfom.
\label{euler-HI}
\end{equation}
(Here and below, unless specified, all integrals are understood over
the domain occupied by the fluid.  An approriate conservative boundary
condition, such as zero normal velocity, is implied.)  In addition to
the explicit integrals \re{euler-HI}, there is also the infinity of
Kelvin invariants,
\begin{equation}
I_\ga=\oint_\ga\bfv\cdot d\bfell\equiv\int_\ga\bfom\cdot d\bfS=\const,
\label{circulation}
\end{equation}
expressing the velocity circulation around (or the vorticity flux
through) any closed contour $\ga(t)$ moving with the fluid velocity
$\bfv$.  Integrals \re{circulation} are {\em implicit\/} in the sense
that their definition involves contours $\ga$ whose motion must be
solved from Eq.~\re{euler}.  Although there is no apparent way of
incorporating integrals like \re{circulation} in a Lyapunov
functional, Arnold \cite{Arnold65} proposed that the conservation of
all vorticity integrals \re{circulation} has the geometrical meaning
of confining the system to an ``iso-vortical sheet'' in the
infinite-dimensional phase space.  Different sets of initial vorticity
integrals specify different sheets such that the whole phase space is
``foliated,'' as if by iso-surfaces of an integral of motion
(Figure~1).
\begin{figure}
\centerline{\psfig{file=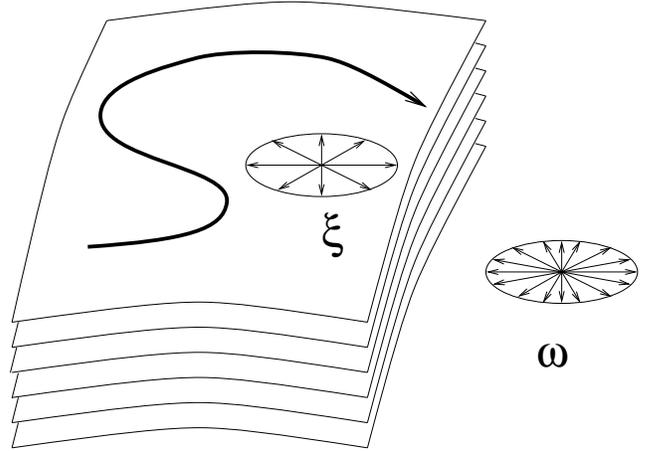,width=3.3in}}
\caption{The infinite-dimensional phase space of all incompressible
fluid flows, $\bfom(\bfx)$, is foliated by iso-vortical invariant sheets
parameterized by the displacement function $\bfxi(\bfx)$.  Each sheet
is an infinite-dimensional subspace of $\bfom$.  The dynamics keeps an
orbit on a sheet.}
\end{figure}
The usefulness of the foliation for stability is due to the local
explicit parameterization of the iso-vortical sheets by an
incompressible ``displacement'' $\bfxi(\bfx)$, such that vorticity
fields sharing the sheet with the reference flow $\bfom_0(\bfx)$ are
written $\bfom=\bfom_0+\de\bfom_0+\frac12\de^2\om_0+\ldots$, where
\begin{equation}
\de\bfom=\bfna\times(\bfxi\times\bfom).
\label{de}
\end{equation}
The iso-vortical variation operator $\de$ derives from the ``modified
dynamics'' $\p_t\bfom=\bfna\times(\p_t\bfxi\times\bfom)$,
$\bfna\cdot\bfxi=0$, in which the vorticity is incompressibly advected
in a way similar to the Euler equation \re{euler}, but by the velocity
field $\p_t\bfxi$ entirely unrelated to the actual flow $\bfv$.  Since
the conservation of the vorticity integrals \re{circulation} is
independent of the relation between $\bfv$ and $\bfom$, Eq.~\re{de}
follows.

Arnold variation \re{de} makes the Hamiltonian $H$ stationary if and
only if the flow $\bfv$ is in equilibrium.  Then the second energy
variation,
\begin{equation}
\de^2H=\int
	(\de\bfv)^2-\bfxi\times\bfom\cdot\bfna\times(\bfxi\times\bfv),
\label{de2}
\end{equation}
if definite for all incompressible $\bfxi$, guarantees that the
equilibrium is stable \cite{Arnold65,Arnold78}.

Given this long introduction, we briefly report on a generalization of
the Arnold method in two important ways.  Firstly, our fluid equations
\re{v}--\re{rh} include compressibility, varying entropy and also
magnetic field, but still no dissipation.  In such a general
formulation, it is difficult to write all integrals generalizing
Eq.~\re{circulation}.  Therefore, and secondly, an analog of the
iso-vortical variation is formally derived from the dynamics, without
regard to either explicit or implicit integrals of motion.  A new
outcome of this procedure is a variational principle for ideal MHD
stability with fluid flow, a long-standing plasma physics problem
\cite{Frieman-Rotenberg-60,Hameiri-Holties-94,Ilgisonis-Pastukhov-96,Berge97}.

We consider the following hydrodynamic equations for an inviscid,
ideally conducting fluid:
\begin{eqnarray}
\rh(\p_t\bfv+\bfv\cdot\bfna\bfv)&=&
	-\bfna p(\rh,s)+\bfj\times\bfB-\rh\bfna\ph,
\label{v}\\
\p_t\bfB&=&\bfna\times(\bfv\times\bfB),
\label{B}\\
\p_ts+\bfv\cdot\bfna s&=&0,
\label{s}\\
\p_t\rh+\bfna\cdot\rh\bfv&=&0.
\label{rh}
\end{eqnarray}
Here $p$ is the fluid pressure, $\rh$ the density, $s$ the entropy,
$\ph$ the external gravitational potential, $\bfB$ the magnetic field,
and $\bfj=\bfna\times\bfB$ the electric current.  The fluid flow
conserves the energy
\begin{equation}
H=\int\rh\left[\frac{\bfv^2}{2}+\ep(\rh,s)+\ph\right]+\frac{\bfB^2}{2},
\label{H}
\end{equation}
where $\ep$ is the specific internal energy defined by the standard
thermodynamic relation 
\begin{equation}
d\ep=T\,ds-p\,d(1/\rh).
\label{dep}
\end{equation}
The varying entropy and the Lorentz force in Eq.~\re{v} break the
``frozen-in'' law for the vorticity $\bfom=\bfna\times\bfv$, and
intead of \re{euler} we now have
\begin{equation}
\p_t\bfom=\bfna\times\left(
	\bfv\times\bfom+\bfj\times\frac\bfB\rh+
	\int^\rh\frac{\p_sp\,d\rh}{\rh^2}\,\bfna s
\right).
\label{om}
\end{equation}

By analogy with the iso-vortical variation \re{de}, one can introduce
modified dynamics for Eqs.~\re{v}--\re{rh} in many different ways
\cite{Gruzinov-variation-95}.  Our choice is dictated by the desire to
have a zero energy variation for equilibrium flows.  Upon a dozen of
attempts, the following procedure works to our satisfaction: We
replace $\bfv\to\p_t\bfxi$ in Eqs.~\re{B}--\re{rh} and \re{om}.  In
\re{om}, we also write $\bfj\to\p_t\bfet$ ($\bfna\cdot\bfet=0$) and
replace the integral by a scalar $\p_t\al$.  The result is {\em the
generalized iso-vortical variation},
\begin{eqnarray}
\de\bfv&=&\bfxi\times\bfom+\bfet\times\frac\bfB\rh+\al\bfna
s+\bfna\be,\quad\bfet=\bfna\times\bfze,
\label{de-v}\\
\de\bfB&=&\bfna\times(\bfxi\times\bfB),\quad
\de s=-\bfxi\cdot\bfna s,\quad
\de\rh=-\bfna\cdot\rh\bfxi,
\label{de-rest}
\end{eqnarray}
which depends on two arbitrary vectors $\bfxi$ and $\bfze$ and two
arbitrary scalars $\al$ and $\be$.

Although variation \re{de-v}--\re{de-rest} conserves magnetic flux and
entropy integrals (which are well known and not written here), it is
unclear what other conservation laws, if any, are accounted for by
this variation.  Nevertheless, the derivation above clearly implies
that the phase-space sheets parameterized by $(\bfxi,\bfze,\al,\be)$
are invariant sheets, which can be interpreted as iso-surfaces of some
integrals of motion and thus used for stability analysis.  In the
limit of zero magnetic field and constant entropy, variation \re{de-v}
reduces to Arnold's iso-vortical variation \re{de}.

The number of arbitrary functions in the variation
\re{de-v}--\re{de-rest} by no accident equals the number of dynamical
equations \re{v}--\re{rh}.  Upon varying the energy \re{H} and using
Eq.~\re{dep}, a few integrations by parts yield
\begin{eqnarray}
\de H&&=\int\bfxi\cdot\left(
	\rh\bfv\cdot\bfna\bfv+\bfna p-\bfj\times\bfB+\rh\bfna\ph
\right)
\nonumber\\
&&-\bfze\cdot\bfna\times(\bfv\times\bfB)+\al\rh\bfv\cdot\bfna s
-\be\bfna\cdot\rh\bfv
\label{de-H}
\end{eqnarray}
By design, the condition that $\de H=0$ for all
$(\bfxi,\bfze,\al,\be)$ is equivalent to an equilibrium solution of
Eqs.~\re{v}--\re{rh}.

The second variation of velocity,
\begin{equation}
\de^2\bfv=\bfxi\times\de\bfom+\bfet\times\de\frac\bfB\rh+\al\bfna\de
s+\bfna\be,
\label{de2-v}
\end{equation}
and similar expressions for $\de^2(\bfB,s,\rh)$ are now used to
calculate the second energy variation:
\begin{eqnarray}
\de^2H=\int&&\de^2\left(\rh\ep+\frac{\bfB^2}2\right)+
\left(\ph+\frac{\bfv^2}2\right)\de^2\rh
\nonumber\\
&&+\rh\,(\de\bfv)^2+\rh\bfv\cdot\de^2\bfv+2\,\de\rh\,\bfv\cdot\de\bfv.
\label{de2-H}
\end{eqnarray}
Equations~\re{de-v}--\re{de-rest} define $\de^2H$ as a functional of
$(\bfxi,\bfze,\al,\be)$.  Two comments regarding the form of $\de^2H$
are in order.

First, the suspicious linear term $\bfna\be$ in the second velocity
variation \re{de2-v} is multiplied by an incompressible $\rh\bfv$ in
Eq.~\re{de2-H} and thus vanishes upon integration by parts.  So,
as it should, $\de^2H$ is a {\em quadratic\/} functional of the
independent variables $(\bfxi,\bfze,\al,\be)$.

Second, the integrand of \re{de2-H} can be written as a quadratic
polynomial of $\al$ with the coefficient $\rh(\bfna s)^2$ in front of
$\al^2$.  Therefore, the definite sign of $\de^2H$ can be only
positive, and, for this, it is necessary and sufficient that the
reduced quadratic form of $(\bfxi,\bfze,\be)$,
\begin{eqnarray}
&&W\equiv\min_\al\de^2H=\int\de^2\left(\rh\ep+\frac{\bfB^2}2\right)+
\left(\ph+\frac{\bfv^2}2\right)\de^2\rh
\nonumber\\
&&-\rh(\bfn\cdot\de''\bfv)^2
+\de'\bfv\cdot(\rh\,\de''\bfv+\bfv\,\de\rh)
+\rh\bfv\cdot\bfet\times\de\frac\bfB\rh,
\label{de2-H1}
\end{eqnarray}
be also positive definite.  Here $\bfn=\bfna s/|\bfna s|$, and the
space-saving notation is introduced,
$$
\de'\bfv\equiv\bfxi\times\bfom+\bfet\times\bfB/\rh+\bfna\be,
\quad\de''\bfv\equiv\de'\bfv+\bfxi\cdot\bfna\bfv-\bfv\cdot\bfna\bfxi.
$$
No further ``simple'' minimization of Eq.~\re{de2-H1} is possible. 

To make Eq.~\re{de2-H1} more explicit, the internal energy variation
can be transformed as
\begin{equation}
\int\de^2(\rh\ep)=\int\bfna\cdot\bfxi
\left(\rh\p_\rh p\,\bfna\cdot\bfxi+\bfxi\cdot\bfna p\right),
\label{de2-internal}
\end{equation}
and the magnetic energy variation
\begin{equation}
\int\de^2\frac{\bfB^2}2=\int\de\bfB\cdot(\de\bfB-\bfxi\times\bfj).
\label{de2-magnetic}
\end{equation}
Also, the second density variation is
$\de^2\rh=\bfna\cdot(\bfxi\bfna\cdot\rh\bfxi)$.

Equation \re{de2-H1} is the main result of this paper.  It can be used
to establish nonlinear stability of general MHD equilibria with fluid
flow, for example numerically, by testing the sign of $W$ for sets of
seven scalar trial functions $(\bfxi,\bfze,\be)$.  Since all known
explicit and implicit integrals of motion are incorporated in our
scheme (the possibly conserved linear and angular momenta amount to
choosing an appropriate frame of reference), {\em we propose that the
sufficient stability criterion $W>0$ is also necessary for the true
nonlinear stability of an ideal MHD equilibrium}.  This conjecture is
also supported by the static limit of zero flow, $\bfv=0$, in which
our variational principle reduces to the sum of Eqs.~\re{de2-internal}
and \re{de2-magnetic}, or the standard MHD energy principle
\cite{BFKK58}, whose violation means a linear instability.  As a
by-prodict, we thus find that the linear stability criterion of
Bernstein {\em et al.}  \cite{BFKK58} for static equilibria is also a
nonlinear stability criterion.  In a general situation with fluid
flow, an indefinite $W$ may not result in an exponential instability,
but rather lead to a slower, nonlinear perturbation growth and
subsequent turbulence.  This scenario will be described elsewhere.

The other two limiting cases we would like to mention are (a) the
hydrostatic equilibrium with $\ph=gz$ and $\bfv=\bfB=0$ and (b) the
incompressible neutral fluid with $\rh=s=\const$ and $\bfB=0$.  For
the former case, the condition $W>0$ yields the well known convective
stability criterion \cite{Lighthill78}: $ds/dz>0$ and $d\rh/dz<0$.  In
the Euler limit, the incompressibility is introduced by letting the
sound speed $c^2=\p_\rh p$ to infinity.  The minimum of
Eq.~\re{de2-internal} then implies $\bfna\cdot\bfxi\to0$ for the
``most dangerous'' perturbations, and further minimization of
\re{de2-H} with respect to $\be$ results in $\bfna\cdot\de\bfv=0$ and
the restricted Arnold criterion that Eq.~\re{de2} be {\em positive\/}
definite.

Thus, in several evident limiting cases, our stability criterion
reduces to the already known results.  However, in the general case of
flow of an inhomogeneous fluid, with or without magnetic field, it
appears new.  The exact mathematical meaning of the generalized
iso-vortical variation and the status of the resulting stability
criterion $W>0$, Eq.~\re{de2-H1}, remain unclear to this author.  For
example, no {\em a priori\/} estimates exist for three-dimensional
hydrodynamic perturbations, unlike those in two dimensions, where all
Casimir integrals are explicit \cite{Arnold78,HMRW85}.  On the
``physical level,'' the sufficient stability criterion $W>0$ looks
rigorous, less so as a necessary one.

{\em Acknowledgments.}  This work was supported by the U.S.\ DOE Grant
No.~DE-FG0388ER53275, the ONR Grant No.~N00014-91-J-1127, and the NSF
Grant PHY94-21318.

\bibliography{../bib/turb,../bib/mbi,../bib/tokamak}

\begin{thebibliography}{10}
\bibitem{Gruzinov-variation-95}
This ambiguity is present already in the Euler equation. For example, one can
  get a different modified dynamics by replacing $\bfom$, and not $\bfv$, by
  $\p_t\bfxi$ in Eq.~\re{euler}. As a result, the variation
  $\de_1\bfom=\bfna\times(\bfv\times\bfxi)$ defines a new foliation of the
  phase space by different invariant sheets. Interestingly, the new variation
  $\de_1$ keeps energy $H$ constant to all orders, but makes the helicity $I$
  stationary only for equilibrium flows and thus allows to study their
  stability by inspecting the second helicity variation $\de_1^2I$. This
  observation was made by A. V. Gruzinov.
\end{thebibliography}

\end{document}